\documentclass{article}

\PassOptionsToPackage{numbers,sort&compress}{natbib}
\usepackage[preprint]{style} 

\usepackage{outlines}

\usepackage[utf8]{inputenc} 
\usepackage[T1]{fontenc}    
\usepackage{hyperref}       
\usepackage{url}            
\usepackage{booktabs}       
\usepackage{amsfonts}       
\usepackage{nicefrac}       
\usepackage{microtype}      
\usepackage{amsmath,amsthm,amsfonts,amssymb,amscd}
\usepackage{mathtools}
\usepackage{lastpage}
\usepackage{enumerate}
\usepackage{fancyhdr}
\usepackage{mathrsfs}
\usepackage[x11names]{xcolor}
\usepackage{graphicx}
\usepackage{listings}
\usepackage{subcaption}
\usepackage[english]{babel}
\usepackage{apxproof}
\usepackage{thmtools}
\usepackage{thm-restate}
\usepackage{enumitem}
\usepackage{float}
\usepackage{bm}
\usepackage{wrapfig}
\usepackage{algorithm}
\usepackage{algpseudocode}

\usepackage{sidecap}
\usepackage{siunitx}
\usepackage{multirow}

\theoremstyle{definition}

\definecolor{myGreen}{rgb}{0,0.5,0}

\title{
Learning, fast and slow 
}

\author{
Markus Meister\\
  Division of Biology and Biological Engineering\\
  Tianqiao and Chrissy Chen Institute for Neuroscience\\
  California Institute of Technology\\
  \texttt{\{meister\}@caltech.edu}
}

\begin{document}

\maketitle

\begin{abstract}
Animals can learn efficiently from a single experience and change their future behavior in response.
However, in other instances, animals learn very slowly, requiring thousands of experiences.
Here I survey tasks involving fast and slow learning and consider some hypotheses for what differentiates the underlying neural mechanisms.
It has been proposed that fast learning relies on neural representations that favor efficient Hebbian modification of synapses.
These efficient representations may be encoded in the genome, resulting in a repertoire of fast learning that differs across species.
Alternatively, the required neural representations may be acquired from experience through a slow process of unsupervised learning from the environment.
\end{abstract}

\section{Introduction}
How animals learn and store information is a central question in brain science. Another area of great interest is how we make decisions between possible actions. The two problems are closely related, in that learning and memory serve the purpose of guiding future decisions. Indeed we generally assess what an animal has learned by tracking how it makes decisions when presented with the identical context before and after learning. 

The present review is about the rate of learning: How much information does the animal extract from its experience and how long does that take? 
 We will see that the rate of learning varies dramatically depending on the kind of task the animal performs. The tasks most popular in laboratory studies of decision-making involve learning at least a factor of 10,000 slower than during natural behaviors. I will discuss possible explanations for this huge discrepancy and some conclusions one may draw for the study of neural mechanisms that implement learning.

\section{How fast do animals learn?}
We will want to compare very different behavioral tasks and species, in the laboratory and in nature, and thus need a quantitative measure of learning that generalizes across these conditions. 
In the most general sense, learning concerns the transfer of information from the environment to the animal, and thus information theory offers a suite of tools needed for measurements and analysis in this domain~\citep{hickRateGainInformation1952,amirValuecomplexityTradeoffExplains2020}. Section \ref{sec:info-methods} lays out the measures used in this article. 

\begin{figure}
  \centering
    \includegraphics[width=0.85\linewidth]{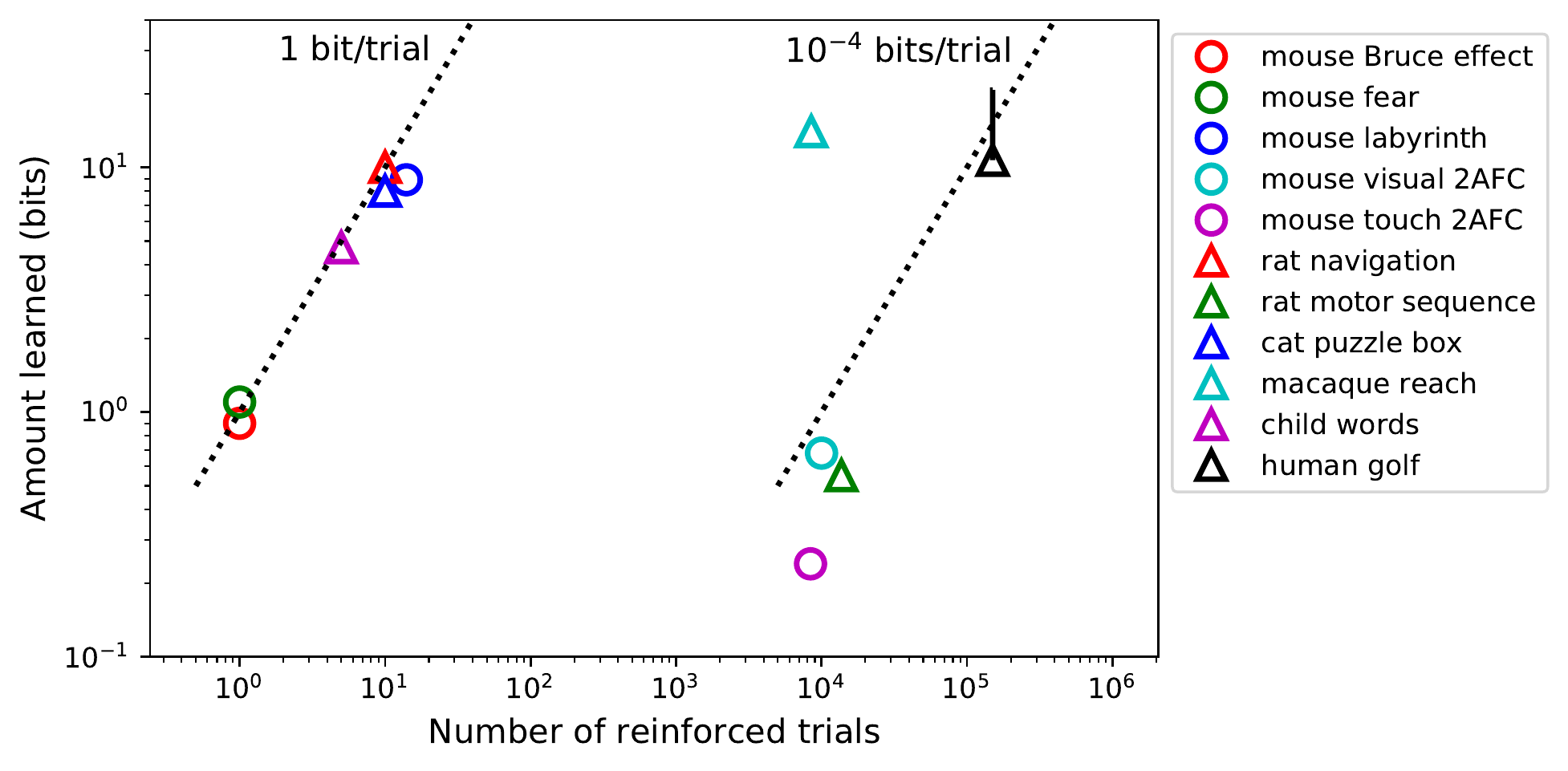}
    \caption{\textbf{Learning rates.} The complexity of various tasks plotted against the number of reinforcement trials required to learn them. Note the learning rates span 4 orders of magnitude. See text and Section \ref{sec:info-methods} for literature and calculations.
    }
\label{fig:learning-rates}
\end{figure}

Figure \ref{fig:learning-rates} summarizes a survey of case studies. It plots the amount of information an animal extracts against the number of experiences required for that learning to take place. Two cautionary notes are in order: First, the broad range of species and activities considered here requires some approximations and assumptions, so the measures here should be considered rough estimates. A more detailed argument may conclude that the number is off by a factor of 2. However, this article is not about factors of 2 but factors of 10,000 or more. Second, the analysis is restricted to the task-relevant information that guides the animal's decision. For example, before a child can learn a new word it must already have acquired a number of generic motor and cognitive skills; here we consider only the process of adding another word to the vocabulary. 

A good laboratory example of "one-shot" learning occurs during fear conditioning in rodents. Here a formerly
innocuous stimulus gets associated with a painful experience, such as electric shock, leading to subsequent avoidance of
the conditioned stimulus~\citep{bourtchuladzeDeficientLongtermMemory1994}. 
Rats and mice will form this association after a single experience
lasting only seconds, and alter their behavior over several hours, with near-perfect recognition of the conditioned stimulus. 
This is an adaptive warning system to deal with life-threatening events, and rapid learning in this case has a clear survival value.
Formally, the resulting behavior maps that stimulus onto one of two actions: freeze vs proceed. 
So the complexity of this task is one bit, representing the maximum amount of information that the animal can transfer from its experience to future behavior (Section \ref{sec:info-methods}).

Another case of learning from single exposure is the Bruce effect: A female mouse learns the smell of her mating partner and uses that to recognize the mate in subsequent encounters. If she meets a strange male that threatens infanticide, she will terminate the pregnancy~\citep{bruceExteroceptiveBlockPregnancy1959,rosserImportanceCentralNoradrenergic1985} about half the time, whereas if she meets her mate again that does not happen. Learning the mate's odor changes her subsequent behavior in response to this particular mouse. Again the complexity of this task is one bit.
Note that the physiological pathways underlying the Bruce effect are innate and do not need to be learned. However, to implement the program the dam must acquire a small amount of olfactory information during the mating experience and retain that through subsequent encounters.

Learning about foods also occurs remarkably fast. For example, laboratory mice with no experience of live prey will overcome their innate aversion to a moving insect after just a handful of encounters~\citep{hoyVisionDrivesAccurate2016}. On the opposite end of the spectrum, animals learn to avoid a food that causes intense nausea after just a single experience~\citep{welzlConditionedTasteAversion2001}.

Of course animals learn more complex tasks as well. For example, mice exploring a labyrinth can learn the shortest route to a reward location that involves many consecutive turning decisions (Figure \ref 
{fig:task-graphs}b). A route that requires 9 bits of information can be learned after about 10 experiences of reward~\citep{rosenbergMiceLabyrinthExhibit2021}. This is also a common theme in many classic studies of rat navigation in mazes. Typically solving the maze involves about 10 correct turns and the learning curves approach perfection after about 10 successful trials~\citep{woodrowProblemGeneralQuantitative1942,munnLearningProcess1950}.  

Another classic branch of animal psychology involves animals escaping from or breaking into a puzzle box~\cite{munnLearningProcess1950}. The animal must discover which among the many possible actions would open the door to freedom. During five minutes of struggle in the box, a cat might attempt several hundred actions. Among those it has to associate one with success, which amounts to learning something in the vicinity of 8 bits. Again, the animals master these tasks after 5-20 successful trials, depending on the intricacy of the lock mechanism~\cite{thorndikeAnimalIntelligenceExperimental1898,danielsBehavioralFlexibilityGeneralist2019}.  

From personal experience we know many cases of fast learning in humans. For example, children and adults will acquire a new word after experiencing it in context only a few times~\citep{careyAcquiringSingleNew1978,bloomHowChildrenLearn2000,yuRapidWordLearning2007}. A typical "Animal menagerie" book for children introduces an animal name for every letter of the alphabet. When a child learns to correctly point to the kangaroo, it has acquired 4.7 bits of task-relevant information (Section \ref{sec:info-methods}). 

All the examples considered so far involve a healthy learning rate of about 1 bit per reinforcement experience (Figure \ref{fig:learning-rates}). Another set of tasks occupies a very different part of the graph, with learning rates of $10^{-4}$ bits per trial or less. These include various trained behaviors that are popular in neural studies of learning and decision-making. 

For example, a long line of research is based on training a macaque monkey to flick its eyes to the left or right when it sees a visual pattern moving to the left or right~\citep{newsomeSelectiveImpairmentMotion1988}. In its simplest version, this is a two-alternative-forced-choice (2AFC) task that requires the association of 2 stimuli with 2 actions (Figure \ref{fig:task-graphs}a). However, the monkey takes many weeks to get halfway good at this task, during which time it performs many thousands of trials with the exact same experience. Another popular task for monkeys is the center-out-reach task: Here the animal has to move its arm to one of eight locations based on a briefly presented visual cue: a task complexity of 24 bits. In a recent study the monkeys achieved 80\% correct performance after 8500 trials, which included careful shaping of the behavior through many intermediate stages~\citep{bergerStandardizedAutomatedTraining2018}.  

In a rodent version of visually driven decision-making, a mouse watches a monitor and handles a steering wheel with its paws. When a stimulus happens on the left or the right side of the screen the animal must turn the wheel accordingly. Again, a simple 2AFC task, but the animal requires about 10,000 trials before its learning saturates, and at that point it performs at 85\% correct~\citep{burgessHighYieldMethodsAccurate2017}. Another task used to study decision-making requires a mouse to sense an object with a single whisker and discriminate two possible object positions~\citep{guoFlowCorticalActivity2014}. Again, the animals must train over many weeks and many thousands of trials to reach a reasonable association of stimuli and actions. In an instance of motor learning, rats were trained to produce two taps at a time interval of 700 ms~\citep{kawaiMotorCortexRequired2015}. After about 14,000 trials they matched this interval to an accuracy of 25\%. Comparing the response distributions before and after training, one concludes that the animals learned about 0.55 bits (Section \ref{sec:info-methods}). 

Humans engage in some slow learning tasks as well, particularly in esoteric domains like art, music, and sports. For example, to become a golf pro takes about 5 years of practice, hitting perhaps 100 drives a day. Top players can land an approach shot within about 5\% of the hole, from varying distances~\citep{ApproachGreenCategories}. Supposing the player could hit twice as far if she wanted and in all possible directions, that amounts to a precision of 11 bits after several years of training (see Section \ref{methods-examples}). Clearly this is another case of ultra-slow learning (Figure \ref{fig:learning-rates}). 

From this survey it emerges that the learning rate spans an enormous range, even within the same species. For example, a mouse exploring a labyrinth extracts about 10 bits of information after receiving 10 rewards, whereas the same mouse strapped into a visual discrimination box will learn a fraction of 1 bit after 5000 rewards. 
Another observation is that tasks phrased in the animal's natural vocabulary -- such as mating, escaping, feeding, or navigation -- get learned much faster than tasks that require abstract or arbitrary associations between contexts and actions. 
The wide gap of 4 log units across learning rates (Figure \ref{fig:learning-rates}) suggests that fast and slow learning are served by very different biological mechanisms. Pursuing that large discrepancy may well reveal something interesting about the brain.

\section{Biological learning is highly constrained by mechanism}

Why are some things so much harder to learn than others? One might have suspected that task complexity is to blame, but that is clearly not the explanation. In Figure \ref{fig:learning-rates} the lowest learning rates appear for certain 2AFC tasks, which have a simple logical structure: two inputs get paired with two outputs. Other 2AFC tasks, with the same logical complexity, are learned much faster. And much more complex tasks, like labyrinths and puzzle boxes, also show high learning rates. So, instead of the structural complexity of the task, we should probably look at the content of what gets learned, namely the specific sets of stimuli and actions that need to be paired.

At the level of animal psychology one might argue that the rodent in a 2AFC box simply doesn't believe what we ask it to learn. The particular contingencies constructed for some of these abstract tasks are highly implausible under natural conditions. Why should the animal believe that touching a pole with a whisker and then stretching the tongue to one side or the other a second later and then receiving a drop of juice are in any way causally connected~\citep{guoFlowCorticalActivity2014}? Among all the possible hypotheses to entertain about the world this one must rank very low in terms of prior probability. Suppose it is 1000 times less plausible than a common hypothesis, like the association of a food odor with impending reward. It is then reasonable, as any Bayesian will attest, that the animal should request 1000-fold stronger evidence for the implausible hypothesis. That could account for the lower learning rates, requiring thousands of repeats of the same implausible coincidence.

In fact, some such Bayesian selection of hypotheses during associative learning seems essential for any productive learning to proceed~\citep{tenenbaumHowGrowMind2011}. Given the huge number of possible pairings that occur between an animal's ongoing sensory streams and its action streams, the brain cannot possibly pursue all those correlations with equal effort. Some choices must be made based on prior plausibility, or associative learning will always chase false leads prompted by accidental coincidences. Those choices will depend on the animal's ecological niche and behavioral needs, so that each biological lineage comes evolutionarily pre-adapted to detect and learn certain contingencies and not others. Those prior beliefs must somehow be encoded in the nervous system. What are some plausible mechanisms for this?

We should consider briefly what is known about the biological mechanisms of associative learning. 
It is generally believed that the substrate of memory resides in the strengths of synapses and the excitability of individual neurons~\citep{kandelMolecularSystemsBiology2014}. 
The process of learning involves changes in those variables -- the synaptic weights and neuronal thresholds -- caused by activity in the network. 
In turn, those changes alter the network's future behavior.
The nature of these activity-dependent changes is constrained by some very restrictive rules.

According to current understanding, the strength of a synapse gets modified depending to the history of activity in the pre- and the post-synaptic neurons~\citep{citriSynapticPlasticityMultiple2008}. 
Of particular importance is the degree of coincident activity, which controls whether the synapse experiences long-term potentiation or depression.
Sometimes the degree of this plasticity is modulated by a third variable, such as a diffuse neuromodulator that may represent the valence of the animal's experience.
In addition there are some non-associative mechanisms, for example a neuron's threshold for firing depends on its own recent activity. 
Synaptic plasticity may appear at multiple pre-synaptic and post-synaptic sites, and it can act on several different timescales~\citep{abbottSynapticComputation2004}. 
However, all these effects are ultimately driven by three variables of neural activity: presynaptic firing, the postsynaptic membrane potential, and (optionally) the concentration of a neuromodulator.

To appreciate just how constraining these rules are, it helps to compare to the learning algorithms used in artificial neural networks (ANNs). In a typical optimization algorithm for an ANN, each synaptic weight gets updated in a way that may depend on all the other weights and all the activities in the network~\citep{Goodfellow-et-al-2016}. Every synapse "sees" every other synapse and all the neurons. In a network the size of cortex, the change in one synapse might depend on the state of a billion other variables. In the biological cortex, it depends on only three. One should expect that such a limited learning rule will impose strong constraints on what can and cannot be learned in biological networks.

\section{Neural representations that favor learning}

If biological learning is fundamentally constrained to sensing the correlations in firing between two neurons, that has strong implications for how events should be represented by neural activity in the brain. In particular, efficient learning benefits from a neural representation that is both high-dimensional and sparse, for the following reasons.

\subsection{High dimensionality}

\begin{figure}
  \centering
    \includegraphics[width=0.5\linewidth]{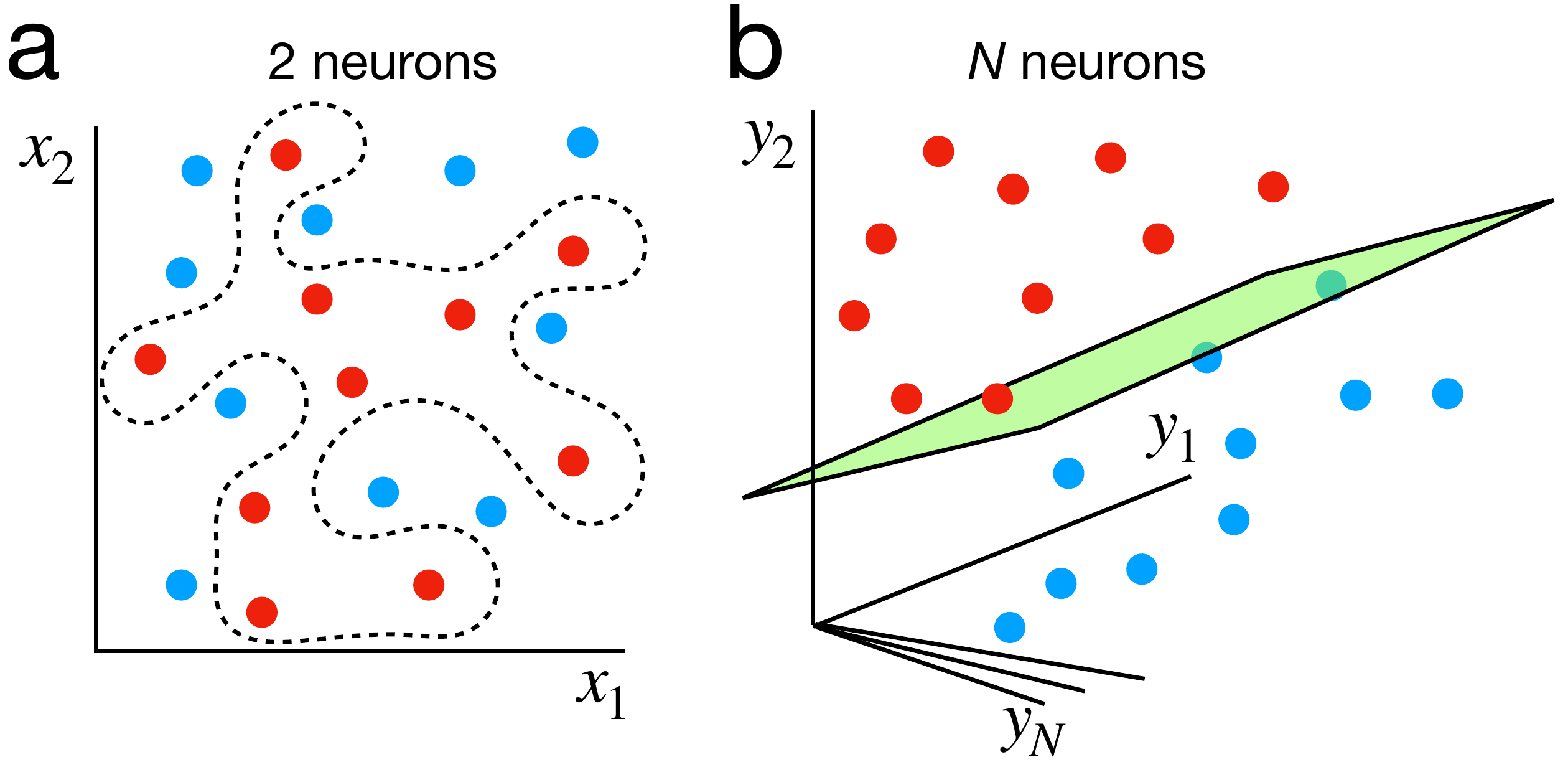}
    \caption{\textbf{Pattern separation in higher dimensions.} \textbf{(a)} Here $m$ different events (dots) are encoded by the firing rate of 2 sensory neurons ($x_1$ and $x_2$). The brain wants to classify those events into good (blue) and bad (red). In the original sensory representation that would require computing the complex region inside the dashed line. \textbf{(b)} After projecting the sensory data into a high-dimensional space, represented by $N>m$ neurons ($y_1,\dots,y_N$), one can generally find a hyperplane (green) such that all the good points are on one side and the bad ones on the other. The projection from sensory signals $x_j$ to the over-complete representation $y_i$ can take the form $y_i = f(\sum_{j=1}^{2} w_{ij} x_j)$ where $w_{ij}$ are random synaptic weights and $f$ is some nonlinear activation function.}
\label{fig:high-dim}
\end{figure}

This means that the representation of events should involve many more neurons than the dimensions in the space of events. \footnote{This is also called an "over-complete" representation.}
To illustrate this, suppose the learning task is a simple classification of events into two groups that are "good" and "bad" based on the animal's experience. In a low-dimensional space, where every event is encoded by a specific combination of neural activities, the good and bad events may be hopelessly interleaved, requiring a difficult computation to separate the two groups (Figure \ref{fig:high-dim}a). 

By contrast, if the events are first mapped into a high-dimensional space, encoded by a large number of neurons, the two groups can be separated on two sides of a hyperplane (Figure \ref{fig:high-dim}b). 
That classification is accomplished by the simplest of neural circuits -- a perceptron -- whose synaptic weights can be learned by a biologically plausible algorithm~\citep{rosenblattPerceptronProbabilisticModel1958,albusTheoryCerebellarFunction1971,dayanTheoreticalNeuroscienceComputational2001}. Note that the high-dimensional projection itself need not be learned: It can use random synaptic weights, but must involve a form of nonlinearity. A number of biological circuits, including the insect mushroom body and the vertebrate cerebellum, are thought to implement such a dimensional expansion that enables pattern separation~\citep{marrTheoryCerebellarCortex1969,albusTheoryCerebellarFunction1971,laurentOlfactoryNetworkDynamics2002,cayco-gajicReevaluatingCircuitMechanisms2019}.

\begin{figure}
  \centering
    \includegraphics[width=0.75\linewidth]{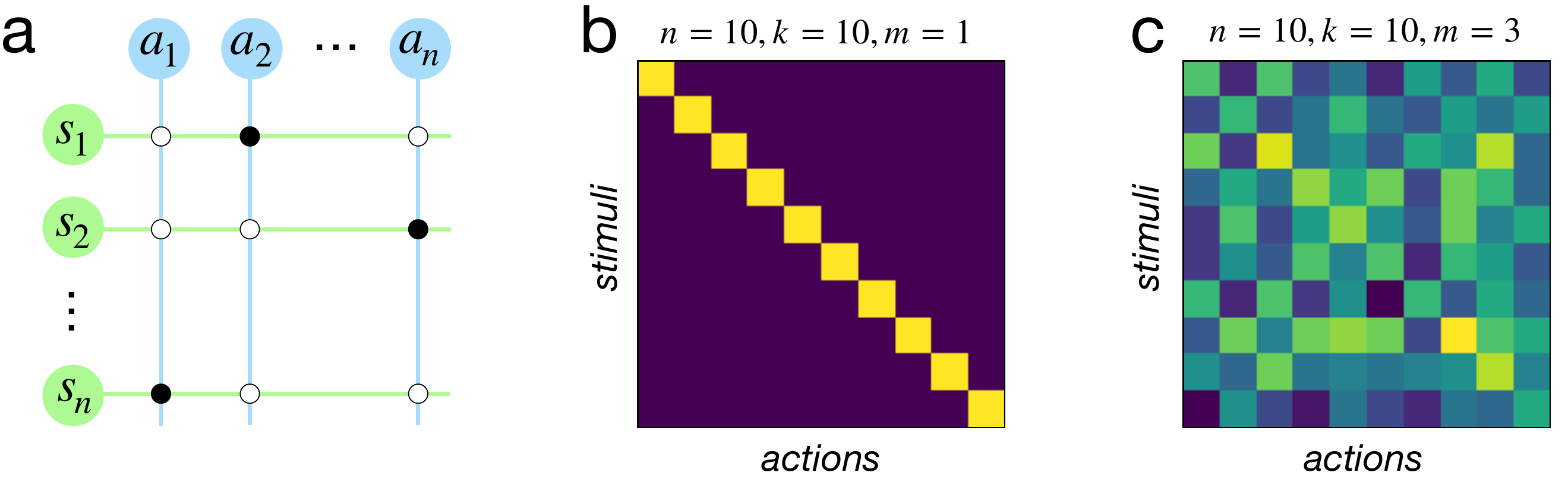}
    \caption{\textbf{Associative learning and sparseness.} \textbf{(a)} A simple network to learn the mapping from a set of states onto a set of actions. Stimuli are represented by $n$ neurons that are either active or inactive, $s_i \in {0,1}$. Similarly actions are represented by $n$ neurons. During learning, the network is exposed to the $k$ desired (state,action) pairs. The synapse from a state neuron to an action neuron is incremented if both pre- and post-synaptic neurons are active. \textbf{(b)} Recall of the stored association: A state vector is presented to the input and the output of the network is compared to the $k$ possible action vectors; this shows the resulting similarity matrix. In this case the state and action vectors each have only $m=1$ of $n=10$ active neurons. The recall of actions associated with each state is perfect. \textbf{(c)} As in panel b, but each vector is represented by 3 active neurons. Note the extensive confusion from the intended mapping of states onto actions.}
\label{fig:assoc-model}
\end{figure}

\subsection{Sparseness}

More general forms of learning go beyond binary classification, and instead require the pair-wise association of one set of events with another set. For example, in learning to navigate a path through a spatial environment or a game, the agent wants to associate the sensory stimuli received at each decision point with the actions that need to be taken there. This kind of pairing can be learned by a network with Hebbian synapses. For this purpose, a suitable representation is not only high-dimensional, but also sparse. In this context, sparseness means that each neuron should be active in only one of the events to be encoded.
\footnote{Sometimes this is called "lifetime sparseness", referring to the fraction of a neuron's life that it is active, as opposed to "population sparseness", which is the fraction of a population active at any one time.}

For illustration, suppose that $N$ stimuli need to be associated with $N$ actions in a one-to-one pairing. The stimuli are presented by $N$ S-neurons in a "one-hot" fashion, meaning each neuron is selective for one of the stimuli (Figure \ref{fig:assoc-model}a). Similarly the actions are represented by $N$ A-neurons in a "one-hot" fashion. The S-neurons connect to the A-neurons via all-to-all Hebbian synapses. This network can learn all $N$ associations after experiencing each pairing just once, because each pairing strengthens exactly one synapse between the two neurons that fire together. From then on, occurrence of each stimulus will cause the activation of the corresponding action neuron (Figure \ref{fig:assoc-model}b). 

By contrast, suppose the stimuli are encoded in a distributed fashion across the S-neurons, and similarly for the actions. As the Hebbian network learns the association of stimuli and actions, each synapse now participates in many of the pairings. This leads to interference between the stored associations, sometimes called "catastrophic forgetting". Thus the capacity of the network is much reduced (Figure \ref{fig:assoc-model}c). Clearly it is advantageous to segregate the information about relevant events into single neurons first, before learning correlations among those events \citep{willshawNonHolographicAssociativeMemory1969,tsodyksAssociativeMemoryNeural1989,palmNeuralAssociativeMemories2013,barlowRedundancyReductionRevisited2001}.

\section{Sparse over-complete neural codes}

One generally finds that the neural representation of relevant events -- such as stimuli and actions -- differs dramatically between the periphery of the nervous system (receptors and muscle fibers) and the central regions. At the periphery, the neural code is strongly determined by the physics of the world (objects causing sounds and images, forces and speeds needed for movement) and the structure of the associated external organs (eye, ear, muscles, and joints). Moving towards the center, the nervous system chooses a different representation that is much more suited to the computational needs of the animal, in particular associative learning~\citep{barlowSingleUnitsSensation1972,olshausenSparseCodingSensory2004}. Here are some examples where we understand the neural code well enough to spot these changes.

Vision starts with the layer of photoreceptors in the retina that use a dense and distributed code. Every object of interest covers many receptors, and the object's identity is hidden in the combinatorial pattern of activity across many neurons. Vice versa, every receptor contributes to many objects. The output of the retina is already substantially transformed: Among retinal ganglion cells there are about 40 different types~\citep{sanesTypesRetinalGanglion2015}, each much more specialized than the photoreceptors~\citep{gollischEyeSmarterScientists2010}. Some of these neurons fire only sparsely under special trigger features, for example differential motion within the image~\citep{zhangMostNumerousGanglion2012}. In primates, the projection from retina through the thalamus to the primary visual cortex involves a huge expansion of neuron numbers by at least a factor of 100. 
\footnote{This is not true in the mouse, where the expansion is only 3-fold.}
At the same time, the neural activity gets even sparser, with the typical cortical neuron firing only 1 spike/s on average~\citep{lennieCostCorticalComputation2003}.
Finally, several synapses later in infero-temporal cortex, neurons are highly selective for those high-level features that are directly relevant to behavior, such as the face of a specific person~\citep{tsaoCorticalRegionConsisting2006}, invariant to position and orientation. 

A similar progression occurs in the olfactory system. Primary receptors respond rather broadly to many molecules. The second-order neurons (projection neurons in insects, mitral cells in vertebrates) are already more selective. Subsequent layers (Kenyon cells, piriform cortex) implement a great expansion, spanning many more neurons than there are primary receptor types. Each of those neurons fires much more sparsely than the receptors, only under certain combinations of odors~\citep{leinwandOlfactoryNetworksSensation2011,turnerOlfactoryRepresentationsDrosophila2008}. 

In the motor system one finds a similar change in neural coding, but in the opposite direction: from sparse representations centrally to dense codes in the periphery. A well-studied example is the motor control of singing in the zebra finch. Signals flow outward from the center to the periphery through a series of nuclei~\citep{feeHypothesisBasalGangliadependent2011}: from a premotor cortical nucleus (called HVC) to a forebrain nucleus (RA) to motor neurons innervating the syrinx. At the level of HVC, the representation is maximally sparse: each neuron fires only at one precise moment during the entire song sequence~\citep{hahnloserUltrasparseCodeUnderlies2002,lynchRhythmicContinuousTimeCoding2016}. Neurons in RA already fire at multiple times during the song, owing to convergence of signals from HVC. Finally, motor neurons in the brain stem receive a massively convergent input from many RA cells. They are active broadly throughout the song, as needed to shape the airway of the syrinx.

\section{How sparse codes may arise}

In all these examples it appears that the central representations of events are eminently suited for rapid associative learning via biological learning rules, in that single neurons are in fact tuned to the specific events of interest. How do these particular transformations arise? There are at least two plausible mechanisms leading to sparse codes: evolution and life-time experience.

Some of the learning-efficient codes are programmed in the genome, presumably as a result of many megayears of trial-and-error experiments. For example, the circuits of the retina are largely hard-wired genetically, with little to no dependence on experience. Retinal ganglion cells are programmed to encode visual changes over space and time, through lateral inhibition in space and delayed inhibition in time. Because visual objects tend to extend over space and time, retinal ganglion cells fire sparsely only at the edges of objects. These physical properties of the world have remained unchanged since the dawn of creation, leaving plenty of time to adapt the neural code. Indeed these coding principles -- namely the focus on temporal or spatial edges -- are remarkably conserved across vertebrate and invertebrate eyes, even though they evolved separately and use different cellular mechanisms.

However, not everything that will be an interesting event for few-shot learning can be anticipated and encoded in genetically programmed circuits. The genome is too small to program even a tiny fraction of the synapses in the brain. So additional learning-efficient codes must be acquired from the animal's lifetime experience. There are many thousands of visual objects that may have behavioral significance in some situations, and it is implausible that special detector circuits evolved for every such object type. Instead, one finds that the sparse code arises only during development, as the animal gains visual experience. For example, in the primary visual cortex the sharp tuning to visual features is strongly dependent on neural activity in early life. 

However, this raises a puzzle: How can a sparse code develop before the animal knows what relationships among visual events it will need to learn some day? One idea is that the visual system learns to develop a sparse code for objects, regardless of any significance of those objects. Each object causes a great deal of redundant structure in the visual image and remains present for some time. Different objects appear and disappear independently of each other. In this way an unsupervised learning system that searches for independent components of an image can eventually learn to represent objects~\citep{dicarloHowDoesBrain2012}. A similar "independent component analysis" is thought to allow us to parse the auditory signal from a cocktail party into streams produced by separate speakers~\citep{haykinCocktailPartyProblem2005}. 

A number of learning mechanisms have been proposed by which the brain might implement independent component analysis~\citep{wiskottSlowFeatureAnalysis2002,foldiakFormingSparseRepresentations1990,hurriSimplecelllikeReceptiveFields2003}. For example the Foldiak network~\citep{foldiakFormingSparseRepresentations1990} combines a feed-forward projection between two layers with recurrent synapses within the output layer. The feed-forward projections obey Hebbian plasticity rules, whereas the recurrent connections are anti-Hebbian. This leads to the emergence in the output layer of pattern detectors for independent components in the input. A succession of such processing stages could then lead to higher levels of abstraction. 

Importantly, though, any such unsupervised online learning is necessarily slow. The statistical regularities that identify the independent events can be assessed and recognized only over many observations. Of course, evolutionary learning takes much longer still.

\section{What distinguishes fast and slow learning?}

We have arrived at two kinds of explanations for the vast difference in learning rates seen across tasks: One is psychological in nature, the other more neural. On the psychological level, animal learning means making some inference about how the world works based on observations. Some of the rules we try to teach animals through reinforcements are so fantastically implausible in the world of that species that they require correspondingly more evidence to overcome that bias. Furthermore, the hypothesis which the experimenter would like the animal to adopt is in constant competition with other accidental contingencies that the animal observes. For example, the mouse might experience an itch on one side of its bum, and three times in a row the water reward appeared on the side of the itch. That correlation will temporarily take over as the favored hypothesis. Even under the carefully controlled conditions of a head-fixed rodent experiment, most of the brain's activity is related to stimuli and actions that have nothing to do with the task at hand~\citep{stringerSpontaneousBehaviorsDrive2019,musallSingletrialNeuralDynamics2019}. Accidental correlations in those signals will overwrite the weak signal from the experimenter's task rules. The task hypothesis may eventually dominate over the competition, but only through a slow process of integration that allows evidence to accumulate over weeks. This can explain both why the performance improves so slowly and why it tends to saturate at a rather modest level~\citep{burgessHighYieldMethodsAccurate2017,guoFlowCorticalActivity2014}.

Human communication plays a special role in overcoming this constraint: A teacher can explain the rules of a card game, and humans will start playing it decently in minutes, whereas monkeys would take months to pick up the rules. Effectively a human teacher can raise the Bayesian prior for an arbitrary abstract hypothesis to 100\% simply by talking about it, and the student will ignore accidental correlations from then on. 

A neural explanation -- which may underlie the psychological one -- is that certain events are predestined for fast learning: If the brain already has a genetically programmed sparse representation for the events that appear in the task, then their association can be formed rapidly. If not, then the brain must first learn a sparse representation of the events in question, which is itself a slow process. Thus the predilection of any given species to learn certain contingencies better than others may just be embodied in the structure of its neural codes.

This interpretation suggests that the abstract 2AFC tasks with minuscule learning rates (Figure \ref{fig:learning-rates}) require a substantial reorganization of the neural code during the training period. Not much empirical information is available on this point, because the animals are usually trained for many weeks before neural recording even begins. However, recent progress in tracking neuronal signals over months~\citep{steinmetzNeuropixelsMiniaturizedHighdensity2021} presents an opportunity to study such changes in neural representation. 

If the hard tasks with low learning rates really require a phase of unsupervised representation learning, one would predict that this can take place even in the absence of reinforcements. As an experimental test of this proposition, one could place the animal into an environment with the same types of abstract stimuli and action choices that appear in the 2AFC task, but without linking them through the task contingencies. Following a few weeks of this, one predicts that (1) the brain should develop concept cells that are specific to key events in the task; and (2) once the contingencies are put in place, the task should be learned much more rapidly. 

Here also, human experiments might present a special opportunity. Recordings from epilepsy patients have revealed neurons that respond sparsely to just one of many possible concepts, like "Bill Clinton"~\citep{quirogaConceptCellsBuilding2012}. If a subject is asked to learn a completely unfamiliar game, one expects to see new concept cells emerge that sparsely represent the key events in the game. Again, a coach can accelerate the learning with verbal explanations. Of course the challenge here is that one can only sample a minute fraction of neurons in the brain, but high-density silicon probes improve the odds significantly.

\section{Do animals benefit from slow learning?}

Given the enormous difference in learning rates between easy and hard tasks (Figure \ref{fig:learning-rates}) one is forced to ask whether the learning of hard tasks actually plays any role in the life of an animal. For a mouse, the 6 weeks of training required to distinguish two abstract stimuli represents a good fraction of the animal's life span. Over that period a wild mouse develops from a newborn to having its own pups. It seems unlikely that the wild mouse uses this time to practice any particular action 10,000 times. Any information in the environment that is relevant to the mouse's fate must be picked up on the first attempt or at most after a few exposures. 

Among nonhuman animals, the zebra finch comes to mind as a possible exception. It spends several weeks of motor practice honing its song, to where it becomes precisely reproducible to millisecond resolution~\citep{tchernichovskiDynamicsVocalImitation2001}. It is thought that the female bird notices this precision and prizes it when selecting a mate, although the actual criteria the female applies remain opaque~\citep{riebelChapterSongFemale2009}. Effectively the male bird's song functions as a mating display~\citep{morrisReproductiveBehaviourZebra1954}, a mental version of the peacock's tail: "See how much brain effort I can waste on this ridiculously hard and pointless task, because everything else comes easy to me." 

We humans have a much longer lifespan. Some of us, at least, can afford to invest the ten thousand hours required to become an expert at something esoteric, like mathematics, or playing the cello, or golf. As a culture we value these feats of slow learning exhibited by our artists, poets, and sports heroes. They stand out in large part because they are so unnatural. They also function as mating displays. So there is a legitimate interest in the mechanisms of slow learning. However, a small short-lived animal like the mouse may not be the best model system, given that it cannot possibly benefit from ultra-slow learning. Of course, six weeks of training on a repetitive task will take its toll, and engrave some kind of circuit in the plastic brain matter of the mouse. But those artificial circuits will have little in common with what the animal uses naturally for learning and decision-making; nor do they have to align with the mechanisms of slow learning in humans.


Nonetheless, prodigious resources are invested today in mouse research focusing on such tasks. The International Brain Laboratory, a consortium of 22 research groups, pursues a mission of understanding everything there is to know about one such behavior: a 2AFC task that requires >10,000 trials of training to get a performance of 85\% correct~\citep{theinternationalbrainlaboratoryStandardizedReproducibleMeasurement2021}. Similarly the Allen Institute, which has mounted an industrial-scale effort towards understanding the mouse brain, uses simple choice tasks in which the animals are trained over more than 10,000 trials before any neural experiment even begins~\citep{groblewskiCharacterizationLearningMotivation2020,garrettExperienceShapesActivity2020}. As a research community we might ask whether it is wise to focus on the lower right portion of Figure~\ref{fig:learning-rates}, where animals learn very little over a very long time. Perhaps an equivalent effort should be targeted at the upper left portion where they learn complex behaviors very quickly.

\appendix
\newpage

\section{Quantitative measures of learning} \label{sec:info-methods}
We are considering a general task defined by a set of states $s_i$ and actions $a_j$ that lead from one state to another. For example, in a simple 2-alternative-forced-choice task to teach a rodent odor classification, the states are given by the two stimuli $s_1 = $ "odor A" and $s_2 = $ "odor B". The actions consist of turning to one of the two reward ports to lick a drop of water: $a_1 = $ "lick left" and $a_2 = $ "lick right". The rodent's task is to learn that $s_1$ should be followed by $a_1$ and $s_2$ by $a_2$. Typically the training regimen involves rewards and punishments that follow correct or incorrect choices of actions. This framework generalizes very broadly to any task that can be described by a graph with states as the nodes, actions as the edges, and rewards distributed along the graph. This includes many natural behaviors like navigation towards a target, and cognitive tasks like games.

\begin{figure}
  \centering
    \includegraphics[width=0.5\linewidth]{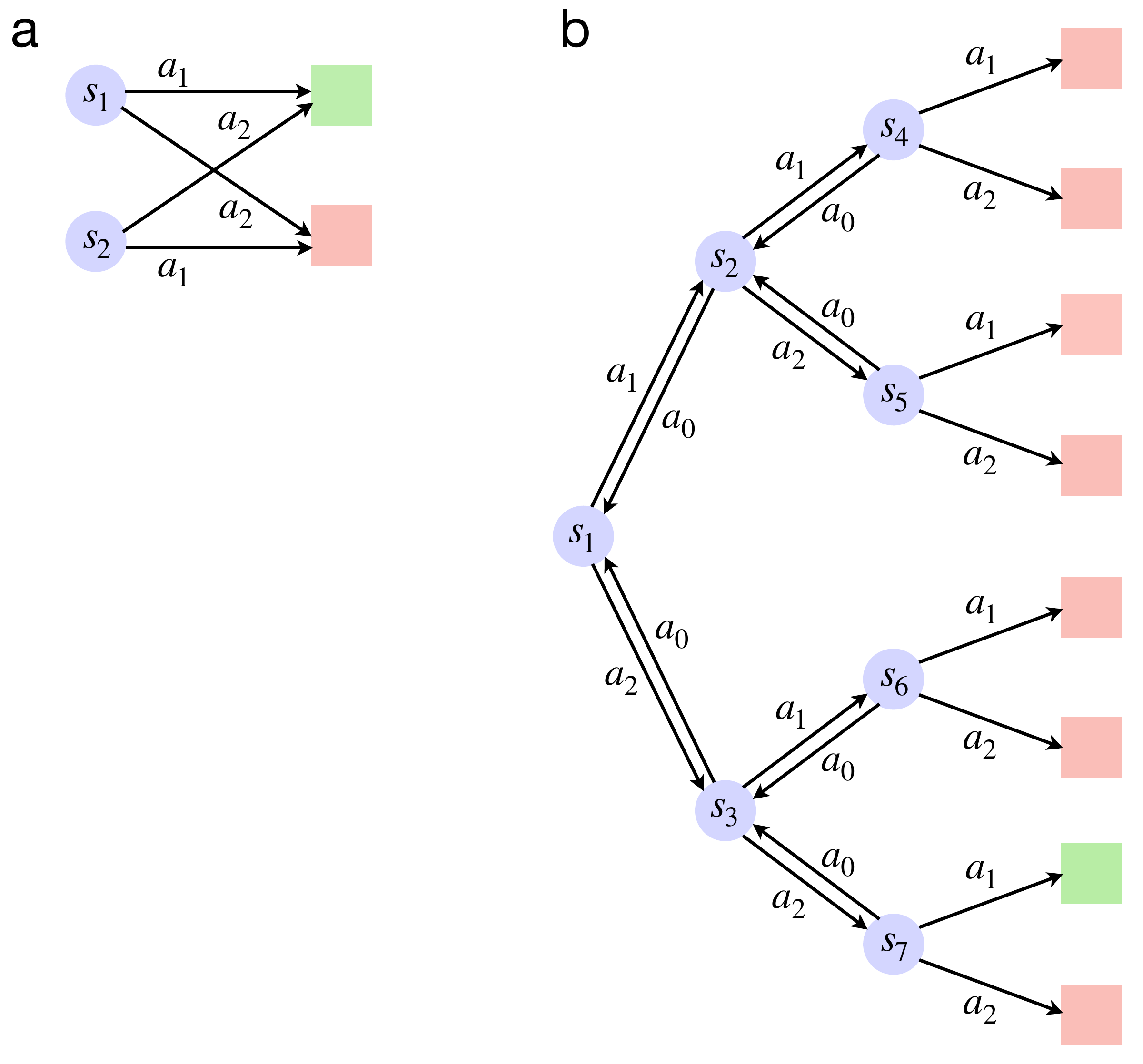}
    \caption{\textbf{Graphs that define behavioral tasks.} \textbf{(a)} A typical 2-alternative-forced-choice task, with states defined by stimuli $s_1$ and $s_2$ and actions of the animal $a_1$ and $a_2$. The square states terminate a trial with reward (green) or no reward (red). \textbf{(b)} A more complex task in which the animal starts in state $s_1$ and then navigates a binary tree by turning left ($a_1$) or right ($a_2$) or backing up to the previous state ($a_0$). Only one of the end points of the tree is rewarded. The task can be implemented as navigation in a binary maze.
    }
\label{fig:task-graphs}
\end{figure}

\subsection{How complex is the task?}
Because learning involves the extraction of information from experience, it seems natural to apply information-theoretic tools to measure learning. Suppose the animal needs to associate $m$ possible states with $n$ possible actions. For each state there is exactly one correct action. Before learning anything, the animal does not know which is the correct pairing of states with actions, and there are $n^m$ possible mappings. Before any learning takes place, the uncertainty about which is the correct mapping can be measured by the entropy of that set, namely 
$$H_{\textrm{bef}}=\log_2 n^m = m \log_2 n$$
For the present purpose we will call that the complexity of the task, 
\begin{equation}
C=H_{\textrm{bef}}
\label{eqn:C-def}
\end{equation}
After learning to perfection, the animal adopts exactly one of these mappings, so the remaining entropy $H_{\textrm{aft}}=0$. The amount the animal learned is the difference between these two entropies 
$$L = H_{\textrm{bef}} - H_{\textrm{aft}}$$
Thus the complexity of the task sets an upper bound on how much the animal can learn.

\subsection{How much did the animal learn?}

Suppose that the animal learns the task incompletely, and fails to associate exactly one action with each state. Instead, from state $s_i$ it produces actions $a_j$ with the conditional probability $p_{\textrm{aft}}(a_j|s_i)$. Then the remaining entropy over all states is
$$H_{\textrm{aft}} = \sum_i \sum_j p_{\textrm{aft}}(a_j|s_i) \log_2 p_{\textrm{aft}}(a_j|s_i)$$
and the amount learned is
\begin{equation}
L = H_{\textrm{bef}} - H_{\textrm{aft}} = \sum_i \sum_j p_{\textrm{aft}}(a_j|s_i) \log_2 \frac{p_{\textrm{aft}}(a_j|s_i)}{p_{\textrm{bef}}(a_j|s_i)} 
\label{eqn:L-def}
\end{equation}
where
$$p_{\textrm{bef}}(a_j|s_i) = \frac{1}{n}$$
is the action probability before learning, uniformly distributed over all possible actions.

\subsection{Examples used in the text} \label{methods-examples}

Perhaps the most elementary learning task is a simple choice between two alternatives, for example when the animal learns which of two response ports is baited with a reward. Here $m=1, n=2$, so the complexity is $C = H_{\textrm{bef}} = 1$ bit. Suppose the animal doesn't learn the task to perfection, but only chooses correctly a fraction $p_{\textrm{aft}}$ of the time. Then the animal has learned 
$$L = H_{\textrm{bef}} - H_{\textrm{aft}} 
= 1 + p_{\textrm{aft}} \log_2 p_{\textrm{aft}} + (1-p_{\textrm{aft}}) \log_2(1-p_{\textrm{aft}}) \textrm{ bits}$$
This scenario applies to fear conditioning, where the sound that was paired with shock needs to be associated with freezing. Responses to other sounds remain the same as before learning. Similarly for the Bruce effect, where the smell of the mate gets associated with carrying the pregnancy to term. The smells of all other males are treated the same. 

The standard 2-alternative-forced-choice task asks the animal to match two possible states (stimuli) with two possible actions. Here $m=2, n=2$, so the complexity is 2 bits. If the animal answers correctly with a probability $p_{\textrm{aft}}$, and that is spread equally over the two conditions, then the amount learned is 
$$L = 2 ( 1 + p_{\textrm{aft}} \log_2 p_{\textrm{aft}} + (1-p_{\textrm{aft}}) \log_2 (1-p_{\textrm{aft}}) ) \textrm{ bits}$$
A monkey performing an 8-direction reaching task has to associate 8 stimuli with 8 movements. If it performs correctly with probability $p_{\textrm{aft}}$ then it has learned 
$$L = 8 ( 3 + (p_{\textrm{aft}} \log_2 p_{\textrm{aft}} + 7 \frac{1-p_{\textrm{aft}}}{7} \log_2 \frac{1-p_{\textrm{aft}}}{7})\textrm{ bits}$$

A child that learns to perfectly match 26 animal names with 26 pictures in a book ($m=n=26$) has learned $26 \log_2 26 \textrm{ bits}$. Every new animal name adds $\log_2 26 = 4.7$ bits.

What if the action distribution is continuous rather than discrete? For example in a study on motor learning~\citep{kawaiMotorCortexRequired2015} rats were trained to produce two taps spaced 700 ms apart. Before training the animals tapped at intervals of 400 ± 200 ms (mean ± SD). After training the distribution was 700 ± 170 ms. How much did the animal learn? Suppose the probability density of the action variable $a$ (here the inter-tap interval) is $p_{\textrm{bef}}(a)$ prior to learning and $p_{\textrm{aft}}(a)$ after. Here we can apply the continuous version of Eqn \ref{eqn:L-def}:
\begin{equation}
L = \int p_{\textrm{aft}}(a) \log_2 \frac{p_{\textrm{aft}}(a)}{p_{\textrm{bef}}(a)} \textrm{d}a
\end{equation}
If the two distributions are roughly Gaussian, this measure amounts to 0.55 bits. This is how much the rat learned about the tapping task it was trained to do.

The case of the golf pro raises some issues. Here the target is the hole on the green, and there are good statistics available on how close the top golfers get to the hole with their approach shots. However we don't have comparable statistics about beginners prior to training; personal experience suggests that their shots are distributed more or less independently of the hole. Instead, here I ask how many different actions the golf pro has available from which she chooses the best. From 100 yards distance she can place the ball into a 5 yard radius. Assuming she could hit the ball to 200 yards, there are 1600 circles of 5 yard radius in that area. So she is able to select correctly one of 1600 possible actions, and $L = H_{\textrm{bef}} - H_{\textrm{aft}} \approx 11$ bits. Clearly this is a rough ball park estimate. Furthermore it supposes that the golf pro practices the same shot over the years. Instead, the ball can lie in different ways relative to the hole, and the player must adjust to each condition. So there are likely multiple states rather than just one and the shot must be learned separately for these conditions. Then the amount learned would increase proportionally. Therefore the estimate in Figure \ref{fig:learning-rates} is indicated as a lower bound. 

\subsection{Why use information theory?}
What recommends these information-theoretic measures of behavior, as compared to any other statistical tool that correlates states with actions? An important desirable for any measure of task complexity is additivity: If a task is composed of two independent subtasks, the complexity of the combined task should be the sum of the individual complexities. The definition for complexity in Eqn \ref{eqn:C-def} satisfies this additivity constraint. For example, this allows us to derive the complexity of the binary decision tree in Figure \ref{fig:task-graphs}b: The shortest route from state $s_1$ to the reward requires one correct 2-way decision followed by two correct 3-way decisions. If the actions "left", "right" and "back" are equally probable to begin with, the complexity is $T = \log_2{2} + 2 \times \log_2{3} = 4.2$ bits. A larger tree with 64 end nodes, as used in a recent experimental study~\citep{rosenbergMiceLabyrinthExhibit2021}, will have task complexity $T = \log_2{2} + 5 \times \log_2{3} = 8.9$ bits.


\section{Annotation of select references} \label{sec:annots}

** Rosenberg et al 2021~\citep{rosenbergMiceLabyrinthExhibit2021}: A study of fast learning in mice during exploration of a labyrinth, combining classic tools of animal psychology with modern measures of behavior.

* Munn 1950~\citep{munnLearningProcess1950}: This book offers a useful overview of classic results in animal learning.

* Palm 2013~\citep{palmNeuralAssociativeMemories2013}: A useful review of associative networks using biological learning rules.

** Olshausen and Field 2004~\citep{olshausenSparseCodingSensory2004}: A cardinal review of sparse coding in many sensory systems.

** Lynch et al 2016~\citep{lynchRhythmicContinuousTimeCoding2016}: A remarkable case of sparse coding in the central representation of a motor program.

* Foldiak 1990~\citep{foldiakFormingSparseRepresentations1990}: An influential paper on unsupervised learning of sparse codes using a simple network with biological learning rules.

* Musall et al 2019~\citep{musallSingletrialNeuralDynamics2019} and Stringer et al 2019~\citep{stringerSpontaneousBehaviorsDrive2019}: Documents how much brain activity is governed by variables unrelated to the task that the experimenters would like to teach the animal.

** IBL 2021~\citep{theinternationalbrainlaboratoryStandardizedReproducibleMeasurement2021}: Report from the International Brain Laboratory on their collective study of a mouse behavior task with ultra-low learning rate.

* Garrett et al 2020~\citep{garrettExperienceShapesActivity2020}: Report from the Allen Institute on a simple binary detection task in mice that gets acquired with an ultra-low learning rate.

\newpage
\section*{References}
\renewcommand{\bibsection}{}\vspace{0em}
\bibliographystyle{unsrtnat}
\bibliography{Fast_Learning}

\newpage
\section*{Code and data}
Code and data supporting this paper can be found at \url{https://github.com/markusmeister/Learning_Fast_And_Slow}.

\section*{Acknowledgements}
MM acknowledges support from NIH (R01 NS111477), the Simons Collaboration on the Global Brain (543015) and the Tianqiao and Chrissy Chen Institute for Neuroscience at Caltech. Thanks to Ralph Adolphs, Pietro Perona, Ueli Rutishauser, Doris Tsao, and Tony Zador for helpful comments and critiques.

\section*{Declaration of interest}
None.

\end{document}